\documentclass{PoS}
\usepackage{bbm}
\usepackage{amsmath}
\usepackage{mathtools}
\usepackage{multirow}
\usepackage{multicol}
\usepackage{graphicx}
\usepackage{xcolor}
\usepackage[export]{adjustbox}
\usepackage{subfig}
\usepackage{tikz}

\DeclareRobustCommand\circled[1]{\tikz[baseline=(char.base)]{\node[shape=circle,draw,inner sep=2pt] (char) {#1};}}

\title{Nuclear Matrix Elements for Neutrinoless Double Beta Decay from Lattice QCD}

\ShortTitle{Nuclear Matrix Elements for $0 \nu \beta \beta$ from Lattice QCD}

\author{William Detmold\\
        Center for Theoretical Physics, Massachusetts Institute of Technology, MA 02139, USA\\
        E-mail: \email{wdetmold@mit.edu}}

\author{\speaker{David Murphy}\\
        Center for Theoretical Physics, Massachusetts Institute of Technology, MA 02139, USA\\
        E-mail: \email{djmurphy@mit.edu}}

\abstract{While neutrino oscillation experiments have demonstrated
that neutrinos have small, nonzero masses, much remains unknown about
their properties and decay modes. One potential decay mode ---
neutrinoless double beta decay ($0 \nu \beta \beta$) --- is a
particularly interesting target of experimental searches, since its
observation would imply that the neutrino is a Majorana particle, demonstrate
that lepton number conservation is violated in nature, and give
further constraints on the neutrino masses and mixing angles. Relating experimental constraints
on $0 \nu \beta \beta$ decay rates to the neutrino masses, however,
requires theoretical input in the form of non-perturbative nuclear
matrix elements which remain difficult to calculate reliably. In this
talk we will discuss progress towards first-principles calculations of
relevant nuclear matrix elements using lattice QCD and effective field
theory techniques, assuming neutrinoless double beta decay mediated by
a light Majorana neutrino. We will show preliminary results for the $\pi^{-} \rightarrow \pi^{+} e^{-} e^{-}$ transition amplitude computed on a $16^{3} \times 32$ domain wall fermion lattice with a pion mass of 420 MeV, and discuss improved methods applicable to general lattice calculations of $0 \nu \beta \beta$ decay amplitudes.}

\FullConference{The 36th Annual International Symposium on Lattice Field Theory - LATTICE2018\\
		22-28 July, 2018\\
		Michigan State University, East Lansing, Michigan, USA.}

\begin{document}

\section{Introduction}

Neutrinoless double beta decay ($0 \nu \beta \beta$), if observed, would provide a wealth of information about the properties of neutrinos --- including resolving the long-standing question of whether they are Majorana or Dirac fermions --- as well as provide an example of a process violating lepton number conservation, and hence contributing to baryogenesis, in nature. While $0 \nu \beta \beta$ has not been observed to date, it is the subject of a large and active experimental search effort, with bounds on the half-lives of relevant nuclei at the level of $T_{1/2}^{0 \nu} \gtrsim 10^{25} - 10^{26}$ yrs \cite{kamland_zen}. Next-generation experiments currently underway are aiming to raise these bounds by an additional one to two orders of magnitude in the near future.

Relating the measured $0 \nu \beta \beta$ decay rate $T_{1/2}^{0 \nu}$ for a particular nucleus to the effective Majorana neutrino mass $m_{\beta \beta} = \left\vert \sum_{k} U_{ek}^{2} m_{k} \right\vert$, where $m_{k}$ are the neutrino eigenstate masses and $U_{ek}$ are elements of the PMNS neutrino mixing matrix, requires theoretical input in the form of a nuclear matrix element $M^{0 \nu}$ describing the non-perturbative, hadronic part of the decay
\begin{equation}
  \left( T_{1/2}^{0 \nu} \right)^{-1} \propto \left\vert m_{\beta \beta} \right\vert^{2} G^{0 \nu} \left\vert M^{0 \nu} \right\vert^{2}.
\end{equation}
Reliably calculating these matrix elements has proven to be a difficult challenge, with predictions for a given nucleus from different nuclear model calculations differing by 100\% or more \cite{nme_calcs}. Improving this situation is crucial for interpreting experimental results as constraints on the parameters of particular models of neutrinoless double beta decay moving forward.

In principle, lattice QCD provides an entirely \textit{ab-initio} method for determining $M^{0 \nu}$. However, in practice, computing matrix elements of the large nuclei relevant to $0 \nu \beta \beta$ searches is well beyond the computational and algorithmic limits of lattice QCD for the foreseeable future. More realistically, one could hope to compute matrix elements of quark-level processes such as $n n \rightarrow p p e e$, and relate these to matrix elements of many-body systems within an effective field theory framework. Another possibility is to compute matrix elements of small nuclei which could be used to probe the systematics of nuclear model calculations by directly comparing lattice and model predictions. First calculations of the long-distance contributions to the neutrinoful double beta decay process $n n \rightarrow p p e e \nu \nu$ and of the short-distance contributions to neutrinoless double beta decay were reported in Refs.~\cite{nplqcd_2vbb} and \cite{0vbb_lattice_sd}, respectively. In this work we discuss first steps toward computing the long-distance contributions to $0 \nu \beta \beta$.

\section{Methodology}

We assume throughout that neutrinoless double beta decay is mediated by the long-distance light Majorana neutrino exchange mechanism. At low energies, and after integrating out the $W$ boson, the underlying Standard Model interaction responsible for this decay is described by the effective electroweak Hamiltonian
\begin{equation}
  \label{eqn:weak_hamiltonian}
  \mathcal{H}_{W} = 2 \sqrt{2} G_{F} V_{ud} \left( \overline{u}_{L} \gamma_{\mu} d_{L} \right) \left( \overline{e}_{L} \gamma_{\mu} \nu_{e L} \right).
\end{equation}
$0 \nu \beta \beta$ is induced at second order in electroweak perturbation theory, leading to the bilocal matrix element \cite{0vbb_review}
\begin{equation}
  \int d^{4} x \, d^{4} y \left\langle f e e \right\vert T \left\{ \mathcal{H}_{W}(x) \mathcal{H}_{W}(y) \right\} \left\vert i \right\rangle = 4 m_{\beta \beta} G_{F}^{2} V_{ud}^{2} \int d^{4} x \, d^{4} y \, H_{\alpha \beta}(x,y) L_{\alpha \beta}(x,y),
  \label{eqn:dbd_me}
\end{equation}
where
\begin{equation}
  L_{\alpha \beta} \equiv \overline{e}_{L}(p_{1}) \gamma_{\alpha} S_{\nu}(x,y) \gamma_{\beta} e_{L}^{C}(p_{2}) e^{- i p_{1} \cdot x} e^{- i p_{2} \cdot y}
\end{equation}
and
\begin{equation}
  H_{\alpha \beta} \equiv \left\langle f \right\vert T \left\{ \overline{u}_{L}(x) \gamma_{\alpha} d_{L}(x) \overline{u}_{L}(y) \gamma_{\beta} d_{L}(y) \right\} \left\vert i \right\rangle
\label{eqn:hadronic_tensor}
\end{equation}
are tensors describing the leptonic and hadronic parts of the decay, respectively, $S_{\nu}(x,y)$ is the neutrino propagator, and $e^{C}_{L} \equiv C \overline{e}_{L}^{\top}$ denotes charge conjugation.

\begin{figure}[!ht]
\centering
  \subfloat[Type \circled{1}]{\includegraphics[width=0.45\linewidth]{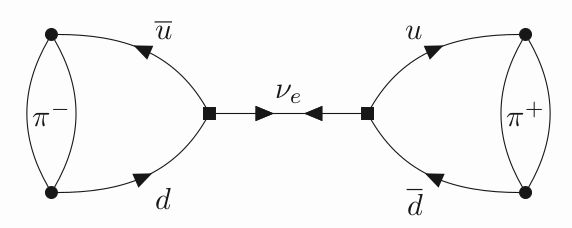}}
\hspace{0.5cm}
  \subfloat[Type \circled{2}]{\includegraphics[width=0.45\linewidth]{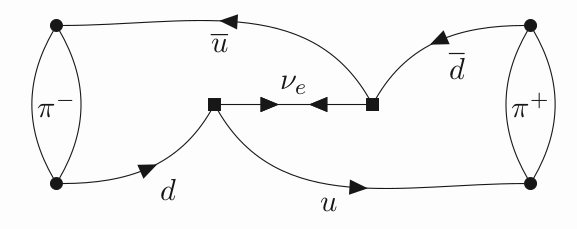}}
\begin{equation}
  \circled{1} = \mathrm{Tr} \left[ S_{u}^{\dagger}(t_{-} \rightarrow x) \gamma_{\alpha} \left( 1 - \gamma_{5} \right) S_{d}(t_{-} \rightarrow x) \right] \cdot \mathrm{Tr} \left[ S_{u}^{\dagger}(t_{+} \rightarrow y) \gamma_{\beta} \left( 1 - \gamma_{5} \right) S_{d}(t_{+} \rightarrow y) \right]
\end{equation}
\begin{equation}
  \circled{2} = \mathrm{Tr} \left[ S_{u}^{\dagger}(t_{+} \rightarrow x) \gamma_{\alpha} \left( 1 - \gamma_{5} \right) S_{d}(t_{-} \rightarrow x) S_{u}^{\dagger}(t_{-} \rightarrow y) \gamma_{\beta} \left( 1 - \gamma_{5} \right) S_{d}(t_{+} \rightarrow y) \right]
\end{equation}
  \caption{Hadronic contractions for the $\pi^{-} \rightarrow \pi^{+} e^{-} e^{-}$ decay. Square dots denote insertions of the effective electroweak Hamiltonian (Eqn.~\eqref{eqn:weak_hamiltonian}).}
  \label{fig:pion_contractions}
\end{figure}

To develop lattice methodology we begin by considering the simplest $0 \nu \beta \beta$ process: $\pi^{-} \rightarrow \pi^{+} e^{-} e^{-}$. Applying Wick's theorem to the hadronic matrix element (Eqn.~\eqref{eqn:hadronic_tensor}) results in two classes of diagrams and four total contractions, depicted in Figure \ref{fig:pion_contractions}.

We denote the time slices of the $\pi^{-}$ source and $\pi^{+}$ sink by $t_{-}$ and $t_{+}$, respectively. The remaining two contractions are obtained by exchanging the locations of the weak current insertions ($x \leftrightarrow y$) and Lorentz indices ($\alpha \leftrightarrow \beta$).

To extract the desired matrix element we employ methods which have been successfully applied to other second-order electroweak processes on the lattice, including the neutrinoful double beta decay process $nn \rightarrow ppee \nu \nu$ \cite{nplqcd_2vbb} and kaon decays \cite{rbcukqcd_rare_kaon, rbcukqcd_deltamk, rbcukqcd_epsilonk}. By inserting a sum over intermediate states $n$ into the bilocal matrix element of Eqn.~\eqref{eqn:dbd_me} it can be shown that the analogous lattice correlation function has the asymptotic time dependence
\begin{equation}
  C_{\pi \rightarrow \pi e e}(t) \propto \sum_{n} \frac{\left\vert Z_{\pi} \right\vert^{2}}{4 m_{\pi}^{2}} \frac{e^{-m_{\pi} t}}{2 E_{n}} \frac{\left\langle \pi e e \right\vert \mathcal{H}_{W} \left\vert n \right\rangle \left\langle n \right\vert \mathcal{H}_{W} \left\vert \pi \right\rangle}{E_{n} - m_{\pi}} \left[ T + \frac{e^{-(E_{n}-m_{\pi}) T} - 1}{E_{n} - m_{\pi}} \right]
  \label{eqn:lattice_bilocal_me}
\end{equation}
for pions at rest, where $T$ is the size of the temporal integration window for the weak current insertions and $t = \vert t_{+} - t_{-} \vert$ is the $\pi^{-} - \pi^{+}$ source-sink separation. In deriving this formula we have assumed that the current insertions are kept sufficiently far from the pion source and sink that potential couplings to excited states may be safely neglected. At large $T$ one can extract the matrix element
\begin{equation}
  M^{0 \nu} = \sum_{n} \frac{\left\langle \pi e e \right\vert \mathcal{H}_{W} \left\vert n \right\rangle \left\langle n \right\vert \mathcal{H}_{W} \left\vert \pi \right\rangle}{E_{n} - m_{\pi}}
  \label{eqn:spectral_me}
\end{equation}
from a linear fit to the $T$ dependence of Eqn.~\eqref{eqn:lattice_bilocal_me}.

In the present context we expect the lowest energy intermediate states to consist of a purely leptonic state $\vert e \overline{\nu}_{e} \rangle$ and a single pion state $\vert \pi e \overline{\nu}_{e} \rangle$, which require special consideration. The $\vert e \overline{\nu}_{e} \rangle$ state contributes a term to Eqn.~\eqref{eqn:lattice_bilocal_me} which grows exponentially as $T \rightarrow \infty$, while, for the $\vert \pi e \overline{\nu}_{e} \rangle$ state, the energy denominator $E_{n} - m_{\pi} \approx m_{e}$ becomes small, potentially contributing a term $\propto T^{2}$. The remaining tower of multi-hadron states have energies $E_{n} > m_{\pi}$ and thus will contribute terms to Eqn.~\eqref{eqn:lattice_bilocal_me} which are asymptotically linear at large $T$.

\section{Pilot Lattice Study of the $\pi^{-} \rightarrow \pi^{+} e^{-} e^{-}$ Decay}

We have performed a pilot calculation using 1000 independent gauge field configurations of the $16^{3} \times 32 \times 16$ domain wall fermion (DWF) ensemble described in Ref.~\cite{16I}. This ensemble has a lattice cutoff of $a^{-1} = 1.6$ GeV and a physical volume of $(2 \, \mathrm{fm})^{3}$, with an unphysically heavy quark mass corresponding to a pion mass of $m_{\pi} = 420$ MeV. We use Coulomb gauge-fixed wall source propagators for the quarks, and a free overlap propagator with an infinite temporal extent for the neutrino. Since performing the full integration over the locations of both weak current insertions is prohibitively expensive, we follow the strategy employed in Refs.~\cite{rbcukqcd_rare_kaon, rbcukqcd_deltamk, rbcukqcd_epsilonk} and treat the weak current insertions asymmetrically: the operator at $x$ is fixed at the (spatial) origin while the operator at $y$ is integrated over the spatial directions. Improved methods which will be used in future lattice calculations are discussed in Section \ref{sec:exact_neutrino_methods}.

\begin{figure}[!ht]
\centering
\subfloat{\includegraphics[width=0.48\linewidth]{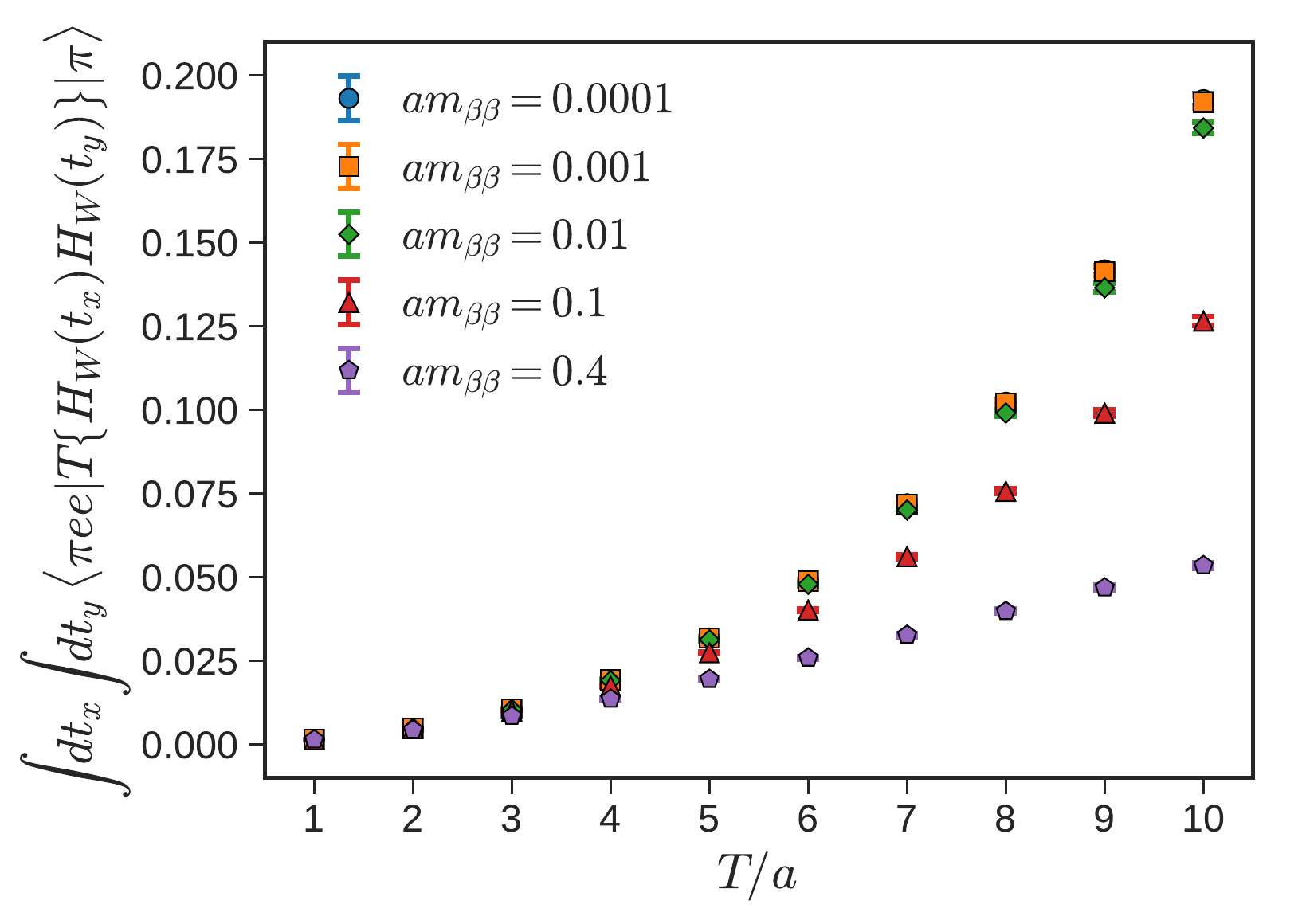}}
\subfloat{\includegraphics[width=0.48\linewidth]{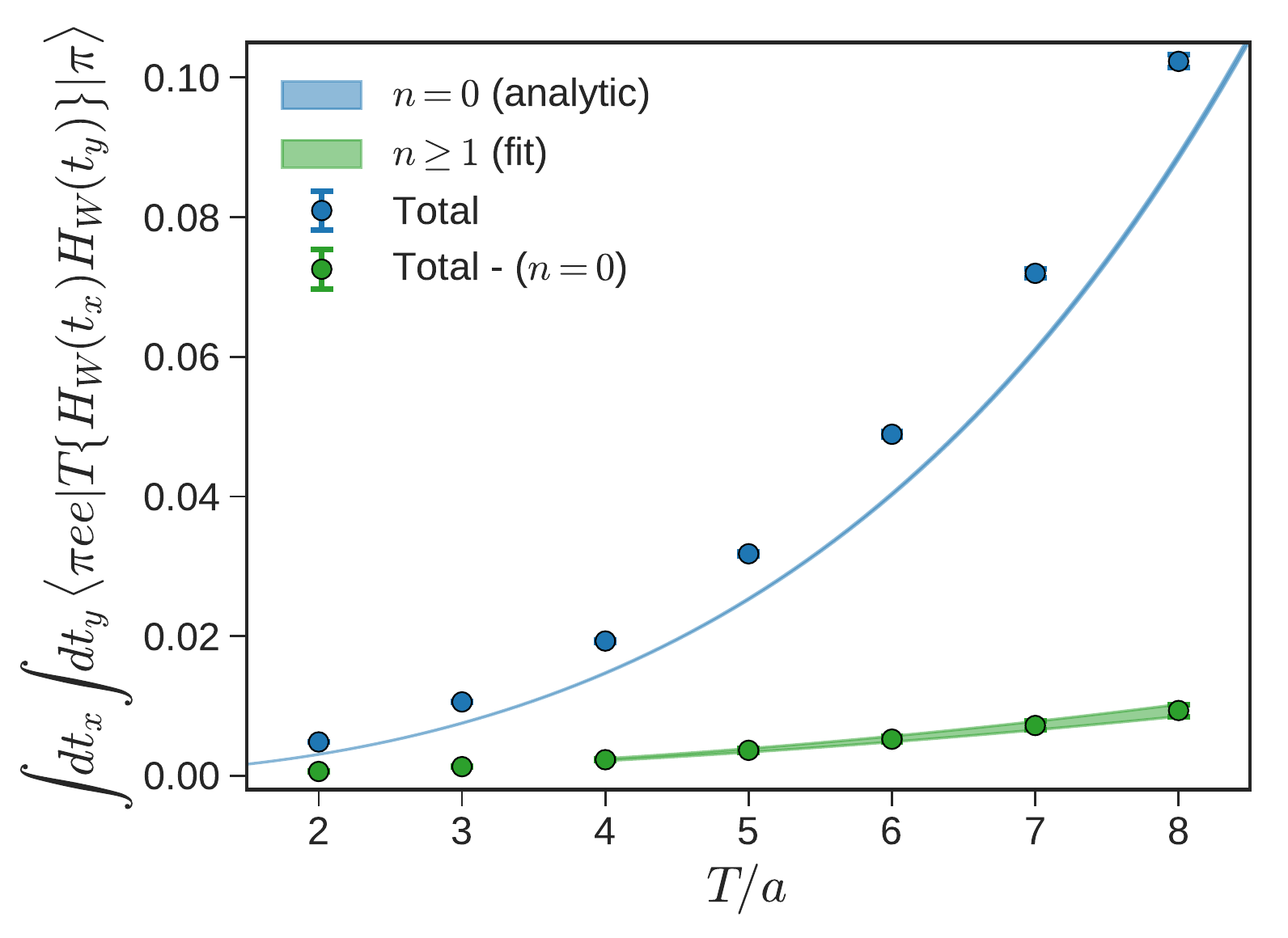}}
  \caption{Left: Preliminary results for integrated bilocal matrix element (Eqn.~\eqref{eqn:dbd_me}) as a function of the temporal extent of the spacetime region used to integrate the weak current insertions, for a variety of neutrino masses given in lattice units. Right: Result for the lightest neutrino mass decomposed into the total, contribution from the $\vert e \overline{\nu}_{e} \rangle$ intermediate state, and the sum over contributions from the single pion and higher energy intermediate states.}
\label{fig:pilot_study}
\end{figure}

In the left panel of Figure \ref{fig:pilot_study} we plot the integrated bilocal matrix element described by Eqns.~\eqref{eqn:dbd_me} and \eqref{eqn:lattice_bilocal_me} as a function of $T$, with the overall factor of $4 m_{\beta \beta} G_{F}^{2} V_{ud}^{2}$ removed, for a wide range of neutrino masses $m_{e} / 3 \lesssim m_{\beta \beta} \lesssim 2 m_{\pi}$. For $m_{\beta \beta} < m_{\pi}$ we observe the expected exponential divergence at large $T$ from the $\vert e \overline{\nu}_{e} \rangle$ intermediate state, as well as the emergence of a consistent $m_{\beta \beta} \rightarrow 0$ limit. We conclude that our calculation is insensitive to the precise choice of $m_{\beta \beta}$ over the range of experimentally relevant neutrino masses. We have also performed the following analysis to extract the matrix element of Eqn.~\eqref{eqn:spectral_me}: we compute the matrix element describing the transition to the vacuum hadronic intermediate state --- $\langle 0 \vert \mathcal{H}_{W} \vert \pi \rangle \propto f_{\pi}$ --- and use this result to analytically construct and subtract the contribution from the $\vert e \overline{\nu}_{e} \rangle$ intermediate state to Eqn.~\eqref{eqn:lattice_bilocal_me}. After performing this subtraction, we then fit a quadratic function in $T$ to the remaining sum over higher intermediate states: from the quadratic term we recover the contribution from the $\vert \pi e \overline{\nu}_{e} \rangle$ state, and from the linear term we recover the sum over the remaining multi-hadron intermediate states. A preliminary analysis is summarized in the right panel of Figure \ref{fig:pilot_study} and in Table \ref{tab:pilot_study}.

\begin{table}[!ht]
\centering
\begin{tabular}{cccc}
\hline
\hline
  \rule{0cm}{0.2cm} & $\vert e \overline{\nu}_{e} \rangle$ & $\vert \pi e \overline{\nu}_{e} \rangle$ & $\vert n e \overline{\nu}_{e} \rangle$ ($n \geq 2$) \\
  $ \left[ \frac{ \langle \pi e e \vert \mathcal{H}_{W} \vert n \rangle \langle n \vert \mathcal{H}_{W} \vert \pi \rangle }{ E_{n} - m_{\pi} } \right] / \left[ \sum_{n} \frac{ \langle \pi e e \vert \mathcal{H}_{W} \vert n \rangle \langle n \vert \mathcal{H}_{W} \vert \pi \rangle }{ E_{n} - m_{\pi} } \right] $ & -0.0082(15) & 1.0082(13) & 0.00009(26) \\
\hline
\hline
\end{tabular}
  \caption{Preliminary results for relative contributions from the hadronic vacuum ($n=0$) and single pion ($n=1$) intermediate states, as well as the sum over all remaining higher energy intermediate states ($n \geq 2$), to the matrix element of Eqn.~\eqref{eqn:spectral_me}.}
\label{tab:pilot_study}
\end{table}

In addition to extracting the matrix element Eqn.~\eqref{eqn:spectral_me}, lattice data for the quark mass dependence of the $\pi^{-} \rightarrow \pi^{+} e^{-} e^{-}$ amplitude can also be matched to the known $\chi$PT amplitude \cite{0vbb_chpt} to extract the next-to leading order low energy constant $g^{\pi \pi}_{\nu}$. First steps in this direction have been performed in Ref.~\cite{xfeng_0vbb_pipiee}, where it was reported that the amplitude is 24\% and 9\% smaller than the leading order $\chi$PT prediction at $m_{\pi} = 420$ MeV and $m_{\pi} = 140$ MeV, respectively. Performing an explicit matching using our results with data at additional pion masses will be the subject of a future study.

\section{Exact Treatment of the Neutrino Propagator}
\label{sec:exact_neutrino_methods}

Lattice QCD calculations of many-body systems are known to suffer from signal-to-noise problems.

In anticipation of future calculations with baryonic and nuclear initial and final states, where we expect such signal-to-noise problems to enter, we have explored methods for performing an exact integration of the matrix element \eqref{eqn:dbd_me} over the spacetime locations of both current insertions; naively one expects an $\mathcal{O}(1/\sqrt{V})$ reduction in the statistical error from making use of the full lattice volume compared to the single sum method of our pilot study. We have also explored directly using the (Euclidean) infinite volume, continuum scalar propagator with a Gaussian UV cutoff for the neutrino,
\begin{equation}
  S_{\Lambda}(x,y) = \int \frac{d^{4} q}{\left( 2 \pi \right)^{4}} \frac{1}{q^{2}} e^{i q \cdot \left( x - y \right)} e^{-q^{2} / \Lambda^{2}} = \frac{1}{4 \pi^{2} \left\vert x - y \right\vert^{2}} \left( 1 - e^{-\frac{\Lambda^{2}}{4} \left\vert x - y \right\vert^{2}} \right),
\end{equation}
which we expect to reduce finite volume effects compared to using a lattice propagator. Here the UV cutoff is required to render the matrix element of Eqn.~\eqref{eqn:dbd_me} finite since the double integration will include contributions where $x \rightarrow y$. We choose $\Lambda = \pi / a$, where $a$ is the lattice spacing, since this choice automatically enforces the removal of the UV cutoff $\Lambda \rightarrow \infty$ in the continuum limit $a \rightarrow 0$ of the lattice calculation.

Implementing the double sum is more difficult, since an explicit $\mathcal{O}(V^{2})$ double integration is prohibitively expensive even for a modestly-sized lattice calculation running on state-of-the-art computational resources. Fortunately, the translational invariance of the neutrino propagator can be exploited to reduce this to $\mathcal{O}(V \log V)$ using the convolution theorem
\begin{equation}
  \int d^{3} x \, d^{3} y \, f_{\alpha}(x) L_{\alpha \beta}( x - y ) g_{\beta}(y) = \int d^{3} x \, f_{\alpha}(x) \left[ \mathcal{F}^{-1} \left\{ \mathcal{F}(L_{\alpha \beta}) \cdot \mathcal{F}(g_{\beta}) \right\} \right](x-y)
  \label{eqn:convolution_theorem}
\end{equation}
and the fast Fourier transform (FFT). For the type 1 diagram, which factors into a product of traces involving only propagators to $x$ or propagators to $y$, the contractions naturally take the form of Eqn.~\eqref{eqn:convolution_theorem}. For the type 2 diagram, which mixes $x$ and $y$, we compute the convolutions of individual spin-color components and reconstruct the trace when we perform the final integration over $x$.

\begin{figure}[!ht]
\centering
  \includegraphics[width=0.55\linewidth]{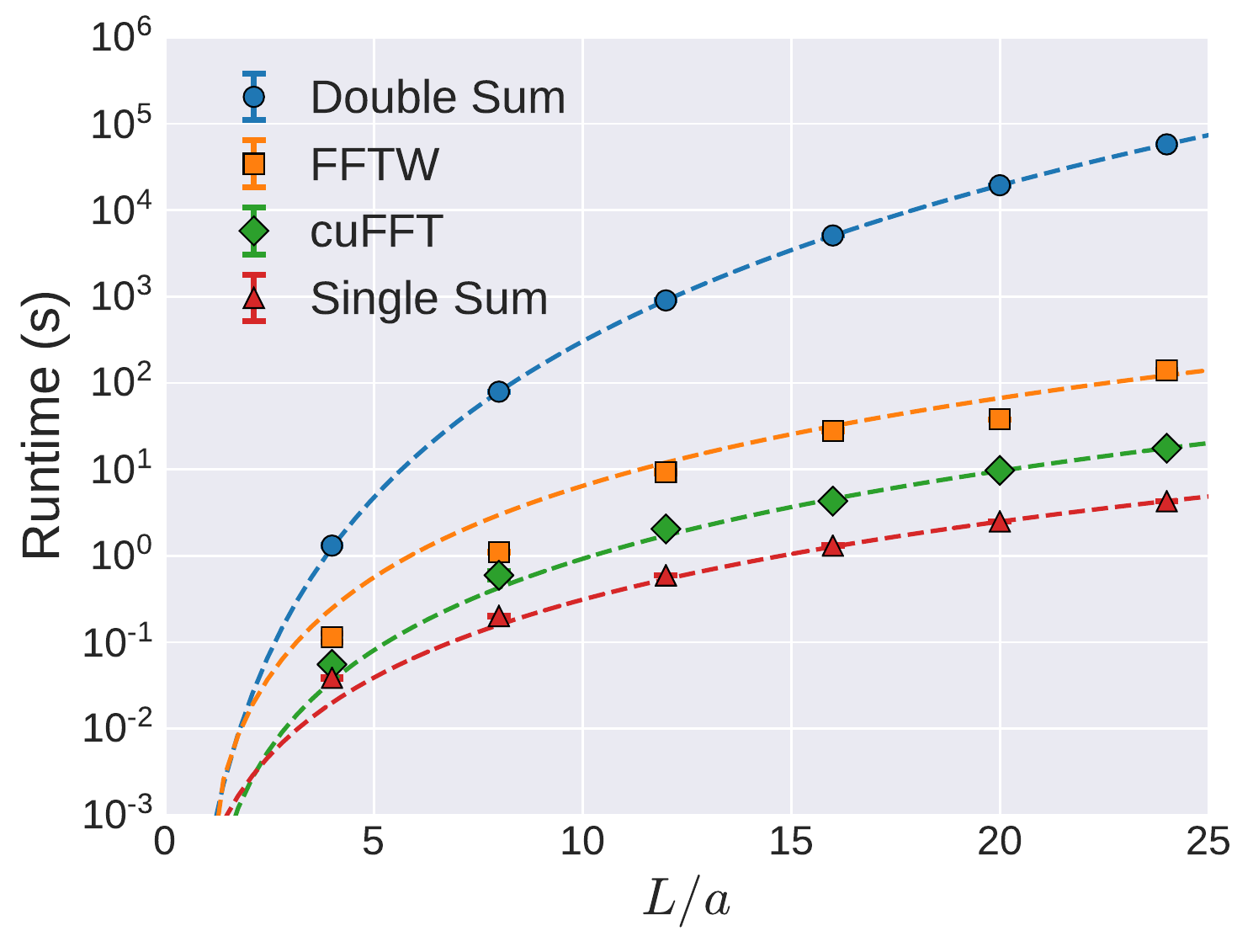}
  \caption{Single-node performance of the FFT-based double summation method described in Section \ref{sec:exact_neutrino_methods} using FFTW running on an Intel Xeon CPU and cuFFT running on an Nvidia GTX 1080 Ti GPU compared to explicit single and double CPU summations for a lattice of spatial length $L$.}
  \label{fig:fft_sum_benchmark}
\end{figure}

In the discrete lattice theory translational invariance implies that the neutrino propagator has a block Toeplitz matrix structure, and algorithms for performing block Toeplitz matrix-vector products via FFTs are well known in the literature. We have chosen to implement an algorithm described in Ref.~\cite{FFT_sum_algo}, which performs the convolution and sum over Lorentz indices at the cost of three one-dimensional FFTs of size $16 \left( 2 L - 1 \right)^{3}$ for a lattice of spatial size $L$. In Figure \ref{fig:fft_sum_benchmark} we benchmark the performance of this algorithm against an explicit double integration, as well as the single integration method used in our pilot study. Runtimes are shown for computing the type 2 contractions integrated over the spatial directions for a single, fixed time ordering of the weak current insertions. We also compare the performance of the OpenMP-threaded FFTW library running on a single Intel Xeon CPU to the performance of the cuFFT library running on an Nvidia GTX 1080 Ti GPU. We find that this strategy is effective in reducing the cost of the double summation to the point that it is feasible for realistic lattice volumes on existing computational resources. We also note a significant performance improvement for the GPU relative to the CPU as the lattice volume grows since large batches of FFTs can be computed in parallel.

\section{Conclusions}

We have performed an exploratory lattice QCD calculation of the $\pi^{-} \rightarrow \pi^{+} e^{-} e^{-}$ transition amplitude on a $16^{3} \times 32 \times 16$ domain wall fermion ensemble, and developed substantially improved methods applicable to general $0 \nu \beta \beta$ decay amplitudes. We are currently using these methods to compute the $\pi^{-} \rightarrow \pi^{+} e^{-} e^{-}$ amplitude on $24^{3} \times 64 \times 16$ domain wall fermion ensembles \cite{24I} at multiple pion masses, and including short-distance contributions \cite{0vbb_lattice_sd} as well as the long-distance contributions described in this work. Analyzing these results and matching them to $\chi$PT, as well as extending our calculations to include baryonic and nuclear initial and final states, will be the subject of future studies.

\section{Acknowledgments}

The authors thank N.~Christ, X.~Feng, R.~Mawhinney, and A.~Pochinsky for helpful discussions which have contributed to this work. W.D.~and D.M.~are partially supported by the U.S.~Department of Energy through Early Career Research Award No.~de-sc0010495 and Grant No.~de-sc0011090, and by the SciDAC4 Grant No.~de-sc0018121. Calculations were performed on the Blue Gene/Q supercomputer at Brookhaven National Lab.

\end{document}